\documentclass[twocolumn,showpacs,preprintnumbers,amsmath,amssymb]{revtex4}
\usepackage{amsmath}
\usepackage{graphicx}% Include figure files
\usepackage{dcolumn}% Align table columns on decimal point
\usepackage{bm}% bold math

\begin{document}

\preprint{APS/123-QED}

\title{Quantifying and identifying the overlapping community structure in networks}

\author{Hua-Wei Shen}
\email{shenhuawei@software.ict.ac.cn}

% \altaffiliation[Also at ]{Institute of Computing Technology, Chinese Academy of Sciences\\}%Lines break automatically or can be forced with \\
\author{Xue-Qi Cheng}%
 \email{cxq@ict.ac.cn}

\author{Jia-Feng Guo}
 \email{guojiafeng@software.ict.ac.cn}

\affiliation{
Institute of Computing Technology, Chinese Academy of Sciences, Beijing, China\\
}%

%\date{\today}% It is always \today, today,
             %  but any date may be explicitly specified

\begin{abstract}
It has been shown that the communities of complex networks often
overlap with each other. However, there is no effective method to
quantify the overlapping community structure. In this paper, we
propose a metric to address this problem. Instead of assuming that
one node can only belong to one community, our metric assumes that a
maximal clique only belongs to one community. In this way, the
overlaps between communities are allowed. To identify the
overlapping community structure, we construct a maximal clique
network from the original network, and prove that the optimization
of our metric on the original network is equivalent to the
optimization of Newman's modularity on the maximal clique network.
Thus the overlapping community structure can be identified through
partitioning the maximal clique network using any modularity
optimization method. The effectiveness of our metric is demonstrated
by extensive tests on both the artificial networks and the real
world networks with known community structure. The application to
the word association network also reproduces excellent results.
\end{abstract}

% PACS, the Physics and Astronomy
                             % Classification Scheme.
%\keywords{Suggested keywords}%Use showkeys class option if keyword
                              %display desired
\maketitle

\section{Introduction}
\label{introduction}

Many complex systems in nature and society can be described in terms
of networks or graphs. The study of networks is crucial to
understand both the structure and the function of these complex
systems~\cite{Albert2002, Newman2003}. A common feature of complex
networks is community structure, i.e., the existence of groups of
nodes such that nodes within a group are much more connected to each
other than to the rest of the network. Communities reflect the
locality of the topological relationships between the elements of
the target systems~\cite{Cheng2009}, and may shed light on the
relation between the structure and the function of complex networks.
Take the World Wide Web as an example, closely hyperlinked web pages
form a community and they often talk about related
topics~\cite{Flake2002}.

\begin{figure}
\vspace{30pt}
\begin{center}
\includegraphics[width=6cm]{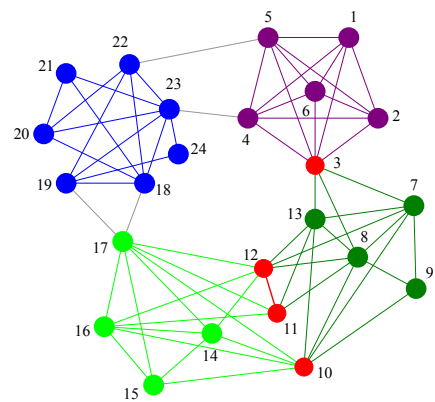}
\end{center}
\vspace{-10pt} \caption{A schematic network with overlapping
community structure. Communities are differentiated by colors and
the overlapping regions are emphasized in red. The edges between
communities are colored in gray.} \label{fig1}
\end{figure}

The identification of community structure has attracted much
attention from various scientific fields. Many methods have been
proposed and applied successfully to some specific complex
networks~\cite{Girvan2002, Newman2004a, Newman2004b, Clauset2004,
Guimera2005, Duch2005, Newman2006, Raghavan2007, Sales-Pardo2007,
Blondel2008}. In order to quantify the community structure of
networks, Newman and Girvan~\cite{Newman2004a} proposed the
modularity as a measure of a partition of network, in which each
node only belongs to one community. The proposal of modularity has
prompted the detection of community structure. However, the
modularity faces several problems. For example, the modularity
suffers a resolution limit problem~\cite{Fortunato2007,Kumpula2007}.
Furthermore, the modularity-based methods cannot tackle overlapping
community structure, in which one node may belong to more than one
community. Figure~\ref{fig1} shows an example network with
overlapping community structure. Intuitively, overlapping community
structure can be represented by a cover of network. A cover of
network is defined as a set of clusters such that each node is
assigned to one or more clusters and no cluster is a proper subset
of any other cluster. As to the network in figure~\ref{fig1}, the
overlapping community structure can be represented by the cover
$\{$$\{1,2,3,4,5,6\}$, $\{3,7,8,9,10,11,12,13\}$,
$\{10,11,12,14,15,16,17\}$,$ \{18,19,20,21,22,23,24\}$$\}$.

Overlapping community structure has been widely
studied~\cite{Palla2005, Baumes2005, Zhang2007, Palla2007,
Farkas2007, Shen2009, Lancichinetti2009a, Nicosia2009, Evans2009}.
In~\cite{Palla2005}, the community structure is uncovered by
$k$-clique percolation and the overlaps between communities are
guaranteed by the fact that one node can participate in more than
one clique. However, the $k$-clique method gives rise to an
uncomplete cover of network, i.e., some nodes may not belong to any
community. In addition, the hierarchical structure can not be
revealed for a given $k$. In~\cite{Nicosia2009}, by introducing the
concept of the belonging coefficients of each node to its
communities, the authors proposed a general framework for extending
the traditional modularity to quantify overlapping community
structure. The method provides a new idea to find overlapping
community structure. However, the physical meaning of the belonging
coefficient lacks a clear explanation. Furthermore, the framework is
hard to extend to large scale networks since it is difficult to find
an efficient algorithm to search the huge solution space. Recently,
Evans et al~\cite{Evans2009} proposed a method to identify the
overlapping community structure by partitioning a line graph
constructed from the original network. This method only allows the
communities to overlap at nodes.

In this paper, a measure for the quality of a cover is proposed to
quantify the overlapping community structure referred as $Q_c$
(quality of a cover). With the measure $Q_c$, the overlapping
community structure can be identified by finding an optimal cover,
i.e., the one with the maximum $Q_c$. The $Q_c$ is based on a
maximal clique view of the original network. A maximal clique is a
clique (i.e. a complete subgraph) which is not a subset of any other
clique in a graph. The maximal clique view is according to a
reasonable assumption that a maximal clique cannot be shared by two
communities due to that it is highly connective. To find an optimal
cover, we construct a maximal clique network from the original
network. We then prove that the optimization of $Q_c$ on the
original network is equivalent to the optimization of the modularity
on the maximal clique network. Thus the overlapping community
structure can be identified through partitioning the maximal clique
network with an efficient modularity optimization algorithm, e.g.,
the fast unfolding algorithm in~\cite{Blondel2008}. The
effectiveness of the measure $Q_c$ is demonstrated by extensive
tests on both the artificial networks and the real world networks
with known community structure and the application to the word
association network.

\section{The quantifying and identifying methods}
\label{method}

In this section, we first propose a measure $Q_c$ to quantify the
overlapping community structure of networks. Then the overlapping
community structure of a network is identified by partitioning a
maximal clique network constructed from the original network using a
modularity optimization algorithm. Finally, some discussions about
our method are given.

\subsection{Quantifying the overlapping community structure}
As mentioned above, the overlapping community structure can be
represented as a cover of network instead of a partition of network.
Therefore, the overlapping community structure can be quantified
through a measure of a cover of network.

As well known, the modularity was used to measure the goodness of a
partition of network. Given an un-weighted, undirected network
$G(E,V)$ and a partition $P$ of the network $G$, the modularity can
be formalized as
\begin{eqnarray}
Q = \frac{1}{L}\sum_{c\in
P}{\sum_{vw}{\delta_{vc}\delta_{wc}\left(A_{vw}-\frac{k_v
k_w}{L}\right)}},\label{eq1}
\end{eqnarray}
where $A$ is the adjacency matrix of the network $G$,
$L=\sum_{vw}A_{vw}$ is the total weight of all the edges, and $k_v =
\sum_{w}A_{vw}$ is the degree of the vertex $v$.

In equation~(\ref{eq1}), $\delta_{vc}$ denotes whether the vertex
$v$ belongs to the community $c$. The value of $\delta_{vc}$ is $1$
when the vertex $v$ belongs to the community $c$ and $0$ otherwise.
For a cover of network, however, a vertex may belong to more than
one community. Thus $\delta_{vc}$ needs to be extended to a
belonging coefficient $\alpha_{vc}$, which reflects how much the
vertex $v$ belongs to the community $c$.

With the belonging coefficient $\alpha_{vc}$, the goodness of a
cover $C$ can be measured by
\begin{eqnarray}
Q_c = \frac{1}{L}\sum_{c\in
C}{\sum_{vw}{\alpha_{vc}\alpha_{wc}\left(A_{vw}-\frac{k_v
k_w}{L}\right)}}.\label{eq2}
\end{eqnarray}

The idea of the belonging coefficient was proposed
in~\cite{Nicosia2009}. Its authors also pointed out that the
belonging coefficient should satisfy a normalization property. This
property is formally written as
\begin{eqnarray}
0\leq \alpha_{vc} \leq 1, \qquad \forall v\in V, \forall c\in
C\label{eq3}
\end{eqnarray}
and
\begin{eqnarray}
\sum_{c\in C}{\alpha_{vc}}=1\label{eq4}.
\end{eqnarray}

Equation~(\ref{eq3}) and equation~(\ref{eq4}) only give the general
constraints on $\alpha_{vc}$, which lead to such a huge solution
space that the enumeration of all the solutions is impractical. To
reduce the solution space and make the problem tractable, we
introduce an additivity property for the belonging coefficient: the
belonging coefficient of a vertex to a community $c$ is the sum of
the belonging coefficients of the vertex to all of $c$'s
sub-communities.

For example, we assume that $C=\{c_1, c_2, \ldots, c_{r-1},$ $c_{r},
\ldots, c_{s}, c_{s+1}, \ldots, c_n\}$ is a cover of the network $G$
and $C'=\{c_1, c_2, \ldots,c_{r-1}, c_{u}, c_{s+1}, \ldots, c_n\}$
is another cover of $G$. The difference between $C'$ and $C$ is that
the community $c_u$ is the union of the communities $c_r, \ldots,
c_s$. The additivity property of belonging coefficient can then be
formally denoted as
\begin{equation}
\alpha_{vc_u} = \sum_{i=r}^{s}{\alpha_{vc_i}}. \label{eq5}
\end{equation}

The belonging coefficient $\alpha_{vc}$ reflects how much a vertex
$v$ belongs to a community $c$. Intuitively, it is proportional to
the total weight of the edges connecting the vertex $v$ to the
vertices in the community $c$, i.e.,
\begin{eqnarray}
\alpha_{vc} \propto \sum_{w\in V(c)}A_{vw},\label{eq6}
\end{eqnarray}
where $V(c)$ denotes the set of vertices belonging to community $c$.
Note that the additivity property of belonging coefficient requires
that communities are disjoint from a proper view of the network.
Therefore, we introduce the maximal clique view to achieve this
purpose. We define $\alpha_{vc}$ as the form
\begin{eqnarray}
\alpha_{vc} = \frac{1}{\alpha_v}\sum_{w\in
V(c)}\frac{O^{c}_{vw}}{O_{vw}}A_{vw},\label{eq7}
\end{eqnarray}
where $O_{vw}$ denotes the number of maximal cliques containing the
edge $(v,w)$ in the whole network, $O^{c}_{vw}$ denotes the number
of maximal cliques containing the edge $(v,w)$ in the community $c$,
and $\alpha_v$ is a normalization term denoted as
\begin{eqnarray}
\alpha_v = \sum_{c\in C}{\sum_{w\in
V(c)}\frac{O^{c}_{vw}}{O_{vw}}A_{vw}}.\label{eq8}
\end{eqnarray}

Obviously, the definition in equation~(\ref{eq7}) satisfies the
normalization property. It also satisfies the additivity property if
we assume that each maximal clique only belongs to one community.
This assumption is reasonable since a maximal clique is highly
connective that any two communities sharing a maximal clique should
be combined into a single one.

With equation~(\ref{eq2}) and equation~(\ref{eq7}), we obtain the
detailed form of $Q_c$ as a measure to the quality of a cover of
network. Note that when a cover degrades to a partition, $Q_c$
becomes the modularity $Q$ in~\cite{Clauset2004} accordingly. In
addition, $Q_c=0$ when all vertices belong to the same community,
and it will be shown later in section~\ref{sec3} that a high value
of $Q_c$ indicates a significant overlapping community structure.

\subsection{Identifying the overlapping community structure}
With the measure $Q_c$, the overlapping community structure of
network can be identified by finding the optimal cover with maximum
$Q_c$. To find the optimal cover, we construct a maximal clique
network from the original network. Then the overlapping community
structure can be identified through partitioning the maximal clique
network.

\subsubsection{Construction of the maximal clique network}
Given an un-weighted, undirected network~$G$, a corresponding
maximal clique network $G'$ can be constructed through the following
method.

\begin{figure*}
\vspace{10pt} \hspace{20pt}
\begin{minipage}[c]{\textwidth}
\begin{center}
\includegraphics[width=14cm]{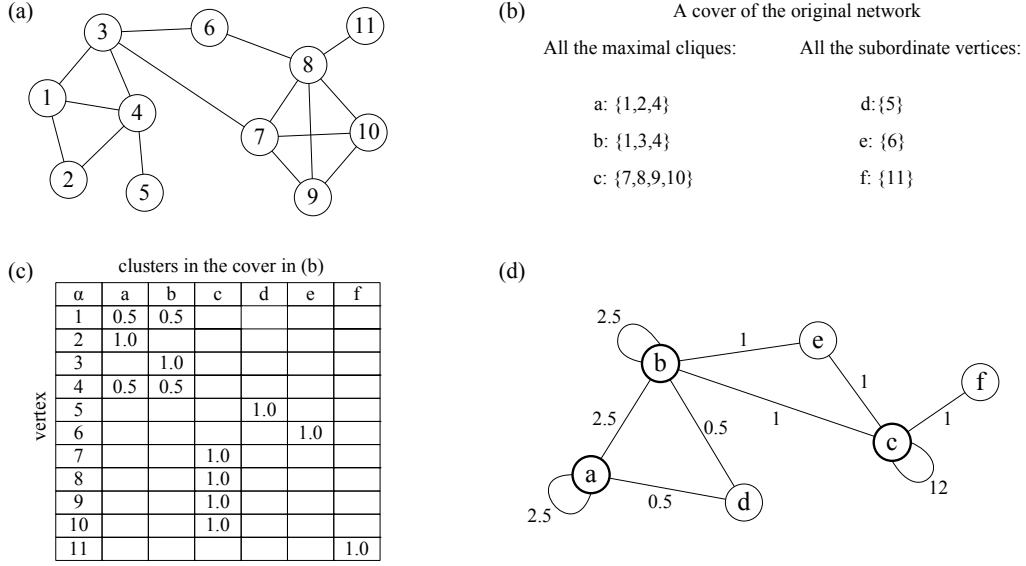}
\end{center}
\end{minipage}
\caption{Illustration for the construction process of the maximal
clique network. (a)The original example network. (b)A cover of the
original network. In this cover, each maximal clique is a cluster
and each subordinate vertex forms a cluster consisting of only one
vertex. (c)The belonging coefficient of each vertex to its
corresponding clusters in the cover. (d) The maximal clique network
constructed from the example network. Here the parameter $k=3$.}
\label{fig2}
\end{figure*}

The maximal clique network $G'$ is constructed by defining its nodes
and edges. We first find out all the maximal cliques in $G$. We can
simply take all these maximal cliques as nodes of $G'$. In practice,
however, we observe that some maximal cliques would not be so highly
connective if their sizes are too small. Such a maximal clique
either lies between different communities (e.g., the maximal cliques
$\{4,23\}$ and $\{5,22\}$ in the network shown in figure~\ref{fig1})
or connects a node to the whole network (e.g., the maximal clique
$\{8,11\}$ in the network shown in figure~\ref{fig2}(a)). To deal
with these small maximal cliques, we introduce a threshold $k$.
Specifically, given the parameter $k$, we only refer to those
maximal cliques with the size no smaller than $k$ as the maximal
cliques, and refer to those with the size smaller than $k$ as
subordinate maximal cliques. We then denote the vertices only
belonging to subordinate maximal cliques as subordinate vertices. In
this way, each maximal clique or subordinate vertex in the original
network $G$ is taken as one node of $G'$.

Note that all the subordinate vertices and the maximal cliques form
a cover $C$ of the original network $G$. For a subordinate vertex
$v$ and a cluster $c$ in the cover $C$, the value of $\alpha_{vc}$
is defined to be $1.0$ when $v$ belongs to the cluster $c$ and $0.0$
otherwise. As to other vertices, $\alpha_{vc}$ can be obtained
according to equation~(\ref{eq7}).

Now we can define the edge of the maximal clique network $G'$ by
defining its adjacency matrix $B$. Let $m_x$ denote the set of the
original network's vertices corresponding to the $x$-th node in
$G'$. The element of $B$ is defined as
\begin{eqnarray}
B_{xy} = \sum_{vw}{\alpha_{vm_x}\alpha_{wm_y}A_{vw}} \label{eq9}
\end{eqnarray}
and the strength (degree) of the $x$-th node
\begin{eqnarray}
s_x = \sum_{y}{B_{xy}}=\sum_{v}{\alpha_{vm_x}k_v}.\label{eq10}
\end{eqnarray}

For clarity, figure~\ref{fig2} illustrates the construction process
of the maximal clique network from an example network with the
parameter $k=3$. Figure~\ref{fig2}(b) shows the subordinate vertices
and the maximal cliques. Each of them becomes a node in the
resulting maximal clique network. For example, the maximal clique
\{1,2,4\} corresponds to the node $a$ and the subordinate
vertex~\{5\} corresponds to the node $d$. Each of these maximal
cliques or subordinate vertices is a cluster in a cover $C$ of the
original network. Their belonging coefficients corresponding to the
cover $C$ are shown in figure~\ref{fig2}(c). According to these
belonging coefficients and equation~(\ref{eq9}), the weight of each
edge of the maximal clique network is obtained. Take the edge
connecting the nodes $a$ and $b$ as an example. As known, the node
$a$ corresponds to the maximal clique \{1,2,4\} and the node $b$
corresponds to the maximal clique \{1,3,4\}. Using the
equation~(\ref{eq9}), the weight of this edge
is~$\alpha_{1a}$$\alpha_{3b}$+$\alpha_{1a}\alpha_{4b}$+$\alpha_{2a}\alpha_{1b}$+$\alpha_{2a}\alpha_{4b}$+$\alpha_{4a}\alpha_{1b}$+$\alpha_{4a}\alpha_{3b}$=$0.5$
+$0.25$+$0.5$+$0.5$+$0.25$+$0.5$=$2.5$.

The constructed maximal clique network is a weighted network though
the original network is un-weighted. The total weight $L'$ of all
the edges in the maximal clique network is equal to the total weight
(number) $L$ of edges in the original network. The proof is
\begin{eqnarray}
L' &=& \sum_{xy}B_{xy} \nonumber \\
   &=& \sum_{xy}{\sum_{vw}{\alpha_{vm_x}\alpha_{wm_y}A_{vw}}} \nonumber \\
   &=& \sum_{vw}{A_{vw}\sum_{x}{\alpha_{vm_x}}\sum_{y}{\alpha_{wm_y}}} \nonumber \\
   &=& \sum_{vw}{A_{vw}} \nonumber \\
   &=& L.\label{eq11}
\end{eqnarray}

Each vertex in the original network corresponds to more than one
node in the maximal clique network. For example, in
figure~\ref{fig2}, the vertex $1$ corresponds to two nodes $a$ and
$b$ in the maximal clique network. Thus, a partition of the maximal
clique network can be mapped to a cover of the original network,
which holds the information about the overlapping community
structure of the original network.

\subsubsection{Finding the overlapping community structure}

\begin{figure}
\vspace{0pt}
%\hspace{40pt}
\begin{minipage}[c]{\columnwidth}
\begin{center}
\includegraphics[width=7cm]{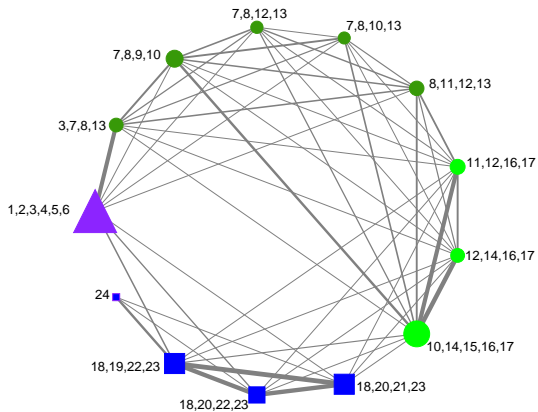}
\end{center}
\end{minipage}
\vspace{-10pt} \caption{The maximal clique network constructed from
the schematic network in figure~\ref{fig1}. The label near each node
shows its corresponding vertices in the original network. The width
of line indicates the weight of the corresponding edge. The
self-loop edge of each node is omitted and its width is reflected by
the volume of the associated circles, squares or triangles. In
addition, the optimal partition of the maximal clique network is
also depicted. The communities in this partition are differentiated
by shapes. Furthermore, the circle-coded community can be
partitioned into two sub-communities. The four communities are shown
in different colors, which are identical to the communities depicted
in figure~\ref{fig1}. Here $k$ is $4$. } \label{fig3}
\end{figure}

Now we investigate the overlapping community structure of the
original network through partitioning its corresponding maximal
clique network. To find the natural partition of a network, the
optimization of modularity is the widely used technique. The
partition with the maximum modularity is regarded as the optimal
partition of network. We employ the algorithm proposed
in~\cite{Blondel2008} to partition our maximal clique network. As an
example, figure~\ref{fig3} shows the partition of a maximal clique
network. Different parts of the partition are differentiated by
shapes or colors.

As mentioned above, each partition of the maximal clique network
corresponds to a cover of the original network and the cover tells
us the overlapping community structure. The key problem lies in that
whether the optimal partition of the maximal clique network
corresponds to the optimal cover of the original network. To answer
this question, we analyze the relation between the modularity of the
maximal clique network and the $Q_c$ of the original network.

Let $\mathcal {P}=\{p_1,p_2,\ldots,p_l\}$ be a partition of the
maximal clique network and $\mathcal{C}=\{c_1,c_2,\ldots,c_l\}$ be
the corresponding cover of the original network. Here, $l$ is the
size of $\mathcal{P}$ or $\mathcal{C}$, i.e., the number of
communities. Using modularity, the quality of the partition
$\mathcal{P}$ can be measured by
\begin{eqnarray}
Q &=& \frac{1}{L'} \sum_{i}{\sum_{x,y\in p_i}{\left(B_{xy}-\frac{s_x
s_y}{L'}\right)}}.\label{eq12}
\end{eqnarray}

Using equations~(\ref{eq9}) and~(\ref{eq10}), we have
\begin{eqnarray}
%Q &=& \frac{1}{L'} \sum_{i}{\sum_{x,y\in
%p_i}{\left(\sum_{vw}{\alpha_{vm_x}\alpha_{wm_y}A_{vw}} - \frac{1}{L'}\sum_{v}{\alpha_{vm_x}k_v}\sum_{w}{\alpha_{wm_y}k_w} \right)}} \nonumber \\
Q &=& \frac{1}{L'} \sum_{i}{\sum_{x,y\in p_i}{\left(\sum_{vw}{\alpha_{vm_x}\alpha_{wm_y}A_{vw}}\right.}} \nonumber \\
&&\qquad \qquad- \left.
\frac{1}{L'}\sum_{v}{\alpha_{vm_x}k_v}\sum_{w}{\alpha_{wm_y}k_w} \right) \nonumber \\
&=& \frac{1}{L'} \sum_{i}{\sum_{x,y\in
p_i}{\sum_{vw}{\alpha_{vm_x}\alpha_{wm_y}\left(A_{vw}-\frac{k_v
  k_w}{L'}\right)}}} \nonumber \\
&=& \frac{1}{L}
\sum_{i}{\sum_{vw}\alpha_{vc_i}\alpha_{wc_i}\left(A_{vw}-\frac{k_v
  k_w}{L}\right)} \nonumber \\
&=& Q_c .\label{eq13}
\end{eqnarray}

Equation~(\ref{eq13}) tells us that the optimization of the $Q_c$ on
the original network is equivalent to the optimization of the
modularity on the maximal clique network. Thus we can find the
optimal cover of the original network by finding the optimal
partition of the corresponding maximal clique network. The optimal
cover reflects the overlapping community structure of the original
network.

\subsection{Discussions}

As to our method, it is important to select an appropriate parameter
$k$. On one hand, the parameter $k$ affects the constituent of the
overlapping regions between communities. According to the definition
to subordinate vertices, they are excluded from the overlapping
regions. Thus the larger the parameter $k$, the less the number of
vertices which can occur in the overlapping regions. When
$k\rightarrow \infty$, the maximal clique network is identical to
the original network and no overlap is identified. On the other
hand, since the subordinate maximal cliques are not so highly
connective, the parameter $k$ should not be too small in practice.
The choice of the parameter $k$ depends on the specific networks.
Observed from many real world networks, the typical value of $k$ is
often between $3$ and $6$. Additionally, as to the networks where
larger cliques are rare, our method is close to the traditional
modularity-based partition methods. In this case, rare overlaps will
be found.

Both the traditional modularity and the $Q_c$ are based on the
significance of link density in communities compared to a null-model
reference network, e.g., the configuration model network. However,
differently from the traditional modularity which requires that each
node can only belong to one community, $Q_c$ requires that each
maximal clique can only belong to one community. In this way, $Q_c$
takes advantage of both the local topological structure (i.e., the
maximal clique) and the global statistical significance of link
density.

The same to the traditional modularity, however, the measure $Q_c$
also suffers the resolution limit problem~\cite{Fortunato2007},
especially when applied to large scale complex networks. Recently,
some methods~\cite{Arenas2008} have been proposed to address the
resolution limit problem of modularity. These methods are also
appropriate to the measure $Q_c$.

Now we turn to the efficiency of our method. It is difficult to give
an analytical form of the computational complexity of our method.
Here we only discuss what influences the efficiency of our method.
Our method consists of three stages, finding out the maximal
cliques, constructing the maximal clique network and partitioning
the maximal clique network. As to the first stage, we need to find
out all the maximal cliques in the network. It is widely believed to
be a non-polynomial problem. However, for real world networks,
finding all the maximal cliques is easy due to the sparseness of
these networks. The computational complexity of the second stage
depends on the number of edges in the original networks. Finally,
the partition stage rests with the number of the maximal cliques and
subordinate vertices. Taken together, our method is very efficient
on real world networks.

In addition, as mentioned above, the overlapping community structure
can be identified by the optimization of $Q_c$. Similarly,
iteratively applying this method to each community, we can
investigate the sub-communities correspondingly. In this way, a
rigid hierarchical relation of overlapping communities can be
identified from the whole network.

\section{Results}
\label{sec3}

In this section, we extensively test our method on the artificial
networks and the real world networks with known community structure.
Then we apply our method to a large real world complex network,
which has been shown to possess overlapping community structure.

\subsection{Tests on artificial networks}

\begin{figure}
\begin{center}
\includegraphics[width=8cm]{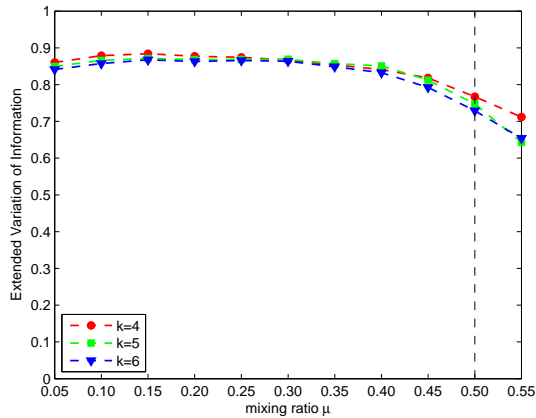}
\end{center}
\caption{Test of our method on the benchmark networks. The parameter
$k$ in the legend corresponds to the parameter $k$ in our method.
The threshold $\mu=0.5$ (dashed vertical line in the figure) marks
the border beyond which communities are no longer defined in the
strong sense~\cite{Radicchi2004}, i.e., such that each node has more
neighbors in its own community than in the others. Each point
corresponds to an average over $100$ graph realization.}
\label{fig4}
\end{figure}

To test our method, we utilize the benchmark proposed
in~\cite{Lancichinetti2009b}. It provides benchmark networks with
heterogenous distributions of node degree and community size. In
addition, it allows for the overlaps between communities. This
benchmark poses a much more severe test to community detection
algorithms than Newman's standard benchmark~\cite{Newman2004a}.
There are many parameters to control the generated networks in this
benchmark, the number of nodes $N$, the average node degree $\langle
k \rangle$, the maximum node degree $maxk$, the mixing ratio $\mu$,
the exponent of the power-law node degree distribution $t1$, the
exponent of the power-law distribution of community size $t2$, the
minimum community size $minc$, the maximum community size $maxc$,
the number of overlapped nodes $on$, and the number of memberships
of each overlapped node $om$. In our tests, we use the default
parameter configuration where $N=1000$, $\langle k \rangle=15$,
$maxk=50$, $t1=2$, $t2=1$, $mic=20$, $maxc=50$, $on=50$ and $om=2$.
By tuning the parameter $\mu$, we test the effectiveness of our
method on the networks with different fuzziness of communities. The
larger the parameter $\mu$, the fuzzier the community structure of
the generated networks is.

To evaluate the effectiveness of an algorithm for the identification
of overlapping community structure, a measure is needed to compare
the cover found by the algorithm with the ground truth.
In~\cite{Lancichinetti2009a}, a measure is proposed to compare two
covers, which is an extension form of \textit{variation of
information}. The more similar two covers are, the higher the value
of the measure is. In this paper, we adopt it to compare the
overlapping community structure found by our method and the known
overlapping community structure in the benchmark networks.

Figure~\ref{fig4} shows the results of our method with $k=4,5,6$ on
the benchmark networks. Our method gives rather good results when
the $\mu$ is smaller than $0.5$. All of the values of the variation
of information are above $0.8$. Note that in these cases,
communities are defined in the strong sense~\cite{Radicchi2004},
i.e., each node has more neighbors in its own community than in the
others. We also test other settings of $k$ which are larger than
$6$, and find similar results.

\subsection{Tests on real world networks}
\begin{figure}
%\vspace{10pt}
\begin{center}
\includegraphics[width=7cm]{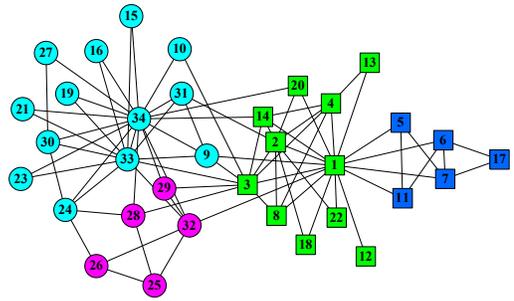}
\end{center}
\vspace{-10pt} \caption{The network of the karate club studied by
Zachary~\cite{Zachary1977}. The real social fission of this network
is represented by two different shapes, circle and square. The
different colors show the partition obtained by our method with the
parameter $k=4$. } \label{fig5}
\end{figure}

Our first real world network for test is Zachary's karate club
network~\cite{Zachary1977}, which is widely used as a benchmark for
the methods of community identification. This network characterizes
the social interactions between the individuals in a karate club at
an American university. A dispute arose between the club's
administrator and its principal karate teacher and as a result the
club eventually split into two smaller clubs, centered around the
administrator and the teacher respectively. The network and its
fission is depicted in figure~\ref{fig5}. The administrator and the
teacher are represented by nodes $1$ and $33$ respectively.

\begin{figure*}
\begin{center}
\includegraphics[width=13cm]{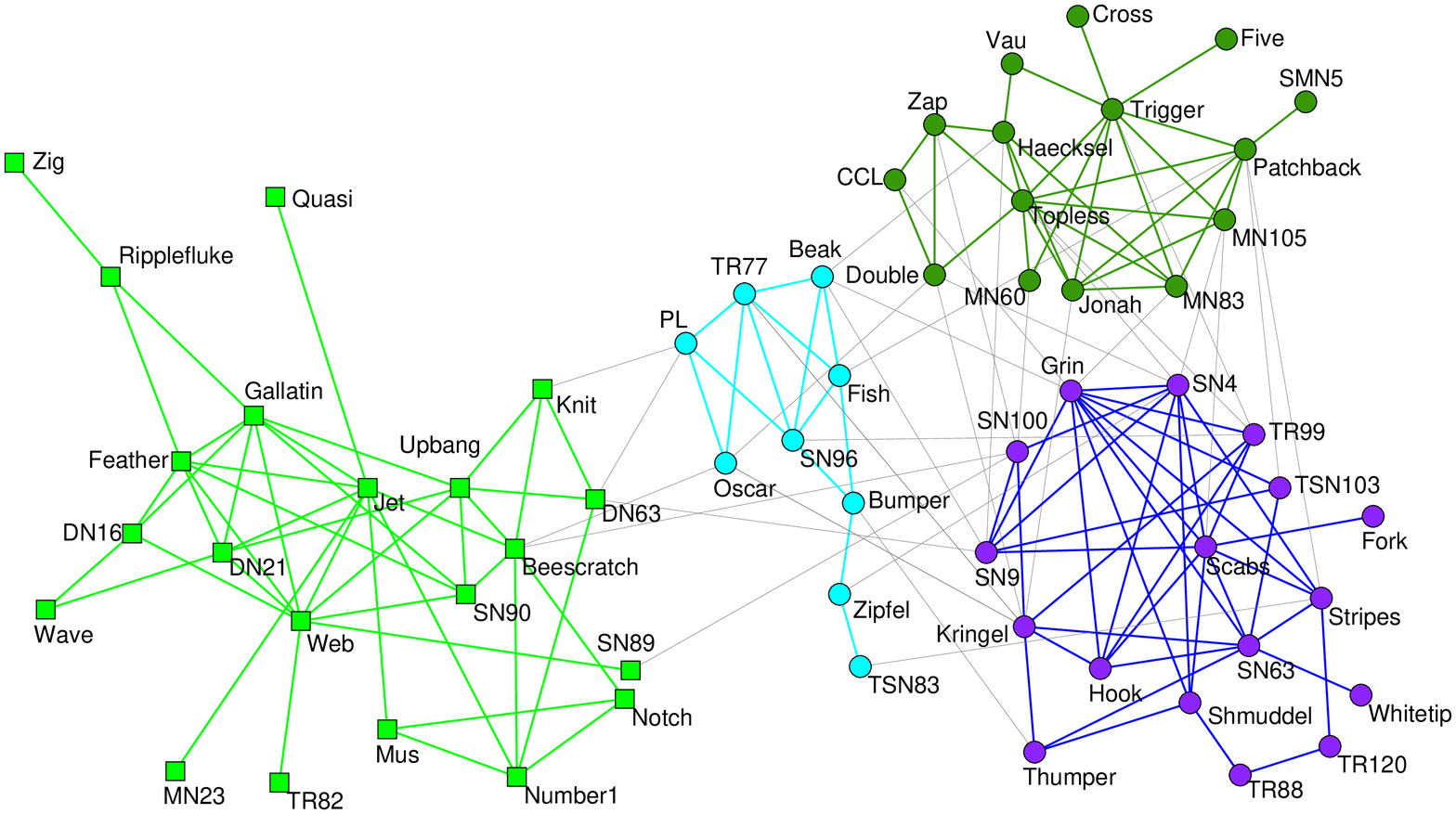}
\end{center}
\caption{The community structure identified by our method from the
network of the bottlenose dolphins of Doubtful Sound. The primary
split of the network is represented by different shapes, square and
circle. The different colors show the partition obtained by our
method with the parameter $k=4$.} \label{fig6}
\end{figure*}

Feeding this network into our method with the parameter $k=4$, we
obtain the result shown in figure~\ref{fig5}. Similar to many
existing community detection methods, our method partitions the
network into four communities. This partition corresponds to the
modularity with the value $0.417$, while the real partition into two
sub-networks has a modularity $0.371$. Actually, no vertex is
misclassified by our method. The real split of the network can be
obtained exactly by pair-wise merge of the four communities found by
our method.

We also note that no overlaps are found when $k=4$. Actually, no
overlaps can be found when $k$ is no smaller than $4$ as to this
network. Overlaps between communities emerge when the parameter $k$
is set to $3$. The value of $Q_c$ corresponding to the resulting
cover is $0.385$ and in total three overlapped communities are found
by our method. They are \{$1$, $5$, $6$, $7$, $11$, $17$\}, \{$1$,
$2$, $3$, $4$, $8$, $9$, $12$, $13$, $14$, $18$, $20$, $22$\} and
\{$3$, $9$, $10$, $15$, $16$, $19$, $21$, $23$, $24$, $25$, $26$,
$27$, $28$, $29$, $30$, $31$, $32$, $33$, $34$\}. The overlapping
regions consist of three vertices, being $1$, $3$ and $9$. Each of
them is shared by two communities. Such vertices are often
misclassified by traditional partition-based community detection
methods. Except the vertices occurring in the overlapping regions,
other vertices reflects the real split of the network.

We also test our method on another real world network, a social
network of $62$ bottlenose dolphins living in Doubtful Sound, New
Zealand. The network was constructed by Lusseau~\cite{Lusseau2003}
with ties between dolphin pairs being established by observation of
statistically significant frequent association. The network splits
naturally into two groups, represented by the squares and circles in
figure~\ref{fig6}.

\begin{table*}
\caption{\label{tab1}The overlapping communities around the word
{\it play}. For each community, a short description is also given.
The overlapped words are emphasized in bold texts.}
%\begin{ruledtabular}
\begin{center}
\begin{tabular}{c|p{2cm}|p{15cm}}
\hline No. & Description & Words in each community \\
\hline 1 & theater & act actor actress bow character cinema curtsey
dance director do drama entertain entertainment film guide involve
juggler lead movie participate perform performance \textbf{play}
portray
producer production program scene screen show sing stage television theater\\
\hline 2 & musical instrument & alto band banjo bass beep blues
brass bugle cello clarinet clef compose concert conductor country
drum faddle fiddle flute guitar harp honk horn instrument jazz
keyboard loud music oboe orchestra piano \textbf{play} rock
saxophone symphony tenor treble trombone trumpet tuba tune
viola violin woodwind\\
\hline 3 & children & adults balls children family friends
\textbf{fun} grown-ups guardians kids love mischief nursery parents
\textbf{play} playground play\rule[-0.2pt]{0.12cm}{0.2pt}dough prank
putty \textbf{toy}
\textbf{toys} tricycle \\
\hline 4 & sports & active arena athlete athletic baseball
basketball
black\rule[-0.2pt]{0.12cm}{0.2pt}and\rule[-0.2pt]{0.12cm}{0.2pt}white
field football \textbf{fun} \textbf{game} illustrated inactive jock
pigskin \textbf{play} recreation referee soccer sports stadium umpire \\
\hline 5 & toys & board boardwalk checkers chess \textbf{fun}
\textbf{game} games monopoly nintendo \textbf{play} plaything
strategy \textbf{toy} \textbf{toys} vcr
video winning yo\rule[2pt]{0.1cm}{0.25pt}yo \\
\hline
\end{tabular}
\end{center}
%\end{ruledtabular}
\end{table*}

By applying our method with $k=4$ to this network, four communities
are obtained, denoted by different colors in figure~\ref{fig6}. The
green community is connected loosely to the other three ones.
Regarding the three circle-denoted communities as a sole community,
it and the green community correspond to the known division observed
by Lusseau~\cite{Lusseau2003}. Furthermore, the three circle-denoted
communities also correspond to a real division among these dolphins.
The further division appears to have some correlation with the
gender of these animals. The blue one consists mainly of females and
the other two almost entirely of males.

Alike to the Zarchay's karate network, the overlaps between
communities cannot be detected when the parameter $k$ is not less
than $4$. When $k=3$, overlaps between the circle-denoted
communities emerge while the green community keeps almost intact.
The $Q_c$ is $0.490$ as to the resulting cover. The vertices
occurring in overlapping regions are $Beak$, $Kringel$, $MN105$,
$Oscar$, $PL$, $SN4$, $SN9$ and $TR99$ among which the vertices
$Beak$ and $Kringel$ are shared by all the three circle-denoted
communities. Again these overlapping vertices are often
misclassified by traditional partition-based methods.

\subsection{Application to the word association network}
Now we apply our method to a large real world complex network,
namely the word association network.

The data set for the word association network is from the demo of
the software {\it CFinder}~\cite{Adamcsek2006}. This network
consists of $7207$ vertices and $31784$ edges, and has been shown to
possess overlapping community structure~\cite{Palla2005}. It is
constructed from the South Florida Free Association norms
list~\cite{Nelson1998}. Initially, the network is a directed,
weighted network. The weight of a directed edge from one word to
another indicates the frequency that the people in the survey
associated the end point of the edge with its start point. These
directed edges were replaced by undirected ones with a weight equal
to the sum of the weights of the corresponding two oppositely
directed edges. Furthermore, the edges with the weight less than
$0.025$ were deleted. In this way, an un-weighted, undirected
network is obtained, and it is the network we deal with.

\begin{figure}
\vspace{10pt}
\begin{center}
\includegraphics[width=7cm]{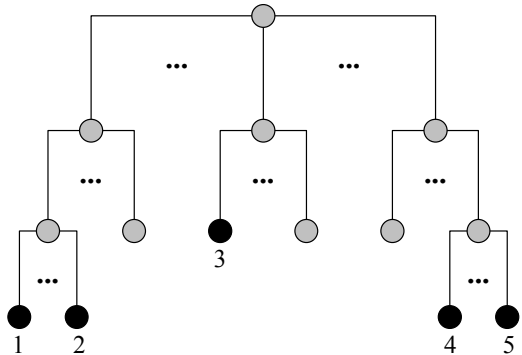}
\end{center}
\vspace{-10pt} \caption{Part of the hierarchy of communities
extracted from the word association network. The dark-filled circles
correspond to the five communities shown in table~\ref{tab1}.}
\label{fig7}
\end{figure}

Applying our method to the word association network, we obtain in
total $20$ communities which overlap with each other. The value of
the corresponding $Q_c$ is as high as $0.503$, indicating a strong
overlapping community structure. The size of these found communities
are very large that there is no specific semantic meaning for each
community. To investigate what is correlated to the overlapping
community structure, we apply our method to these communities
iteratively and a hierarchy of overlapping communities is obtained.
We find that the sub-communities have certain correlation with
semantic meaning of words. As an example, table~\ref{tab1} shows us
the communities around the word $play$. The five overlapping
communities represent different meanings of the word $play$,
respectively related to {\it theater}, {\it musical instruments},
{\it children}, {\it sports} and {\it toys}. Except the
common-shared word {\it play}, four other words are shared by some
of these communities. They are {\it fun}, {\it game}, {\it toy} and
{\it toys}. The overlap between these communities characterizes the
direct, local relationship between them through sharing members.
However, the extent of closeness between communities is sometimes
reflected by the indirect, global relationship between them. One of
this kind of relationship is the ``genealogical'' relationship
between communities, which can be illustrated by the hierarchy of
overlapping communities. Figure~\ref{fig7} is an example for
hierarchy of communities. As shown in figure~\ref{fig7}, the
communities $1$ and $2$ are in the same branch of the hierarchy,
indicating that the meanings represented by them are closer. This
can be validated by examining the words contained in these two
communities. Similarly, the communities $4$ and $5$ are also closely
related. However, the distance between the communities $3$ and $5$
is larger although they share as many as $4$ words. The overlaps
between communities and the hierarchy of these communities provide
us a more complete understanding to the relationship between
communities.

\section{Conclusions}
This paper focuses on the problem of quantifying and identifying the
overlapping community structure of networks. There are two main
contributions. Firstly, a measure $Q_c$ for the quality of a cover
of network is proposed to quantify the overlapping community
structure. The effectiveness of the measure $Q_c$ is demonstrated by
the experimental results that networks with significant overlapping
community structure have a cover with a high $Q_c$. Secondly, a
maximal clique network is constructed from the original network, and
then the overlapping community structure can be identified using any
modularity optimization method on the maximal clique network.

The $Q_c$ is an extension of traditional modularity with the
consideration that the maximal clique instead of a single node can
only belong to one community. In this way, $Q_c$ takes advantage of
both the local topological structure (i.e., the maximal clique) and
the global statistical significance of link density compared with a
null-model reference network. In addition, $Q_c$ can be naturally
used to simultaneously identify the overlapping and hierarchical
community structure of networks. Such a method is helpful to more
completely understand the functional and structural properties of
networks.

As the further work, we will consider the generalization to the
weighted and/or directed networks. It is also an interesting problem
about the selection of the parameter $k$ in our method. We will
further investigate how to determine an appropriate $k$ for a given
network later.

\begin{acknowledgements}
This work was funded by the National Natural Science Foundation of
China under grant number $60873245$, the National High-Tech R\&D
Program of China (the $863$ program) under grant number
$2006AA01Z452$, and the National Basic Research Program of China
(the $973$ program) under grant number $2004CB318109$. The authors
gratefully acknowledge S.~Fortunato and A.~Lancichinetti for
providing the test benchmark and useful discussions on it. The
authors thank Mao-Bin Hu for helpful suggestions.

\end{acknowledgements}

\end{document}